\documentclass[prl,twocolumn,superscriptaddress,showpacs,preprintnumbers,amsmath,amssymb,floats]{revtex4}

\usepackage{txfonts}
\usepackage{amssymb}
\usepackage{graphicx}
\usepackage{amsmath}
\usepackage{amssymb}
\usepackage{ulem}

\begin{document}

\title{A  unifying  phase diagram with correlation-driven  superconductor-to-insulator transition for the 122$^{*}$ series of iron-chalcogenides}

\author{X. H. Niu}
\affiliation{State Key Laboratory of Surface Physics, Department of Physics,  and Advanced Materials Laboratory, Fudan University, Shanghai 200433, People's Republic of China}
\affiliation{Collaborative Innovation Center of Advanced Microstructures, Nanjing 210093, People's Republic of China}
\author{S. D. Chen}
\affiliation{State Key Laboratory of Surface Physics, Department of Physics,  and Advanced Materials Laboratory, Fudan University, Shanghai 200433, People's Republic of China}
\author{J. Jiang}
\author{Z. R. Ye}
\author{T. L. Yu}
\author{D. F. Xu}
\author{M. Xu}
\author{Y. Feng}
\author{Y. J. Yan}
\author{B. P. Xie}
\author{J. Zhao}
\affiliation{State Key Laboratory of Surface Physics, Department of Physics,  and Advanced Materials Laboratory, Fudan University, Shanghai 200433, People's Republic of China}
\affiliation{Collaborative Innovation Center of Advanced Microstructures, Nanjing 210093, People's Republic of China}
\author{D. C. Gu}
\affiliation{Institute of Physics, Chinese Academy of Sciences, Beijing 100190, People's Republic of China}
\author{L. L. Sun}
\affiliation{Institute of Physics, Chinese Academy of Sciences, Beijing 100190, People's Republic of China}
\affiliation{Collaborative Innovation Center of Quantum Matter, Beijing 100190, People's Republic of China}
\author{Qianhui Mao}
\author{Hangdong Wang}
\affiliation{Department of Physics,  Zhejiang University, Hangzhou, 310027, People's Republic of China}
\author{Minghu Fang}
\affiliation{Department of Physics,  Zhejiang University, Hangzhou, 310027, People's Republic of China}
\affiliation{Collaborative Innovation Center of Advanced Microstructures, Nanjing 210093, People's Republic of China}
\author{C. J. Zhang}
\affiliation{High Magnetic Field Laboratory, Chinese Academy of Sciences and University of Science and Technology of China, Hefei 230026, People's Republic of China}
\affiliation{Collaborative Innovation Center of Advanced Microstructures, Nanjing 210093, People's Republic of China}
\author{J. P. Hu}
\affiliation{Institute of Physics, Chinese Academy of Sciences, Beijing 100190, People's Republic of China}
\affiliation{Collaborative Innovation Center of Quantum Matter, Beijing 100190, People's Republic of China}
\author{Z. Sun}\email{sunzhe@gmail.com}
\affiliation{National Synchrotron Radiation Laboratory, University of Science and Technology of China, Hefei, Anhui 230029, People's Republic of China}
\affiliation{Collaborative Innovation Center of Advanced Microstructures, Nanjing 210093, People's Republic of China}
\author{D. L. Feng}\email{dlfeng@fudan.edu.cn}
\affiliation{State Key Laboratory of Surface Physics, Department of Physics, and Advanced Materials Laboratory, Fudan University, Shanghai 200433, People's Republic of China}
\affiliation{Collaborative Innovation Center of Advanced Microstructures, Nanjing 210093, People's Republic of China}

\begin{abstract}

The 122$^{*}$ series of iron-chalcogenide superconductors,  for example K$_x$Fe$_{2-y}$Se$_{2}$, only possesses electron Fermi pockets. Their distinctive electronic structure challenges the picture built upon iron pnictide superconductors, where  both electron and hole Fermi pockets coexist. However, partly due to the intrinsic phase separation in this family of compounds, many aspects of their behavior remain elusive.  In particular, the evolution of the 122$^{*}$ series of iron-chalcogenides with chemical substitution  still lacks a microscopic and unified interpretation.
Using angle-resolved photoemission spectroscopy, we studied a major fraction  of 122$^{*}$ iron-chalcogenides, including the  isovalently `doped' K$_x$Fe$_{2-y}$Se$_{2-z}$S$_z$,  Rb$_x$Fe$_{2-y}$Se$_{2-z}$Te$_z$ and (Tl,K)$_x$Fe$_{2-y}$Se$_{2-z}$S$_z$. We found that the bandwidths of the low energy Fe \textit{3d} bands  in these materials depend on doping;  and more crucially, as the bandwidth decreases, the ground state evolves from a metal to a superconductor, and eventually to an insulator, 
yet the Fermi surface  in the metallic phases is unaffected by the isovalent dopants. 
Moreover, the correlation-driven insulator found here with small band filling may  be a novel insulating phase.
Our study  shows that almost all the known 122$^{*}$-series iron chalcogenides can be understood {\it via} one unifying phase diagram which implies that moderate correlation strength is beneficial for the superconductivity. 

\end{abstract}


\maketitle

\section{I. Introduction}

The iron-based superconductors (FeHTSs)  can essentially be classifed into two large families based on their electronic structure. The first family is comprised of  iron pnictides and  Fe(Te,Se) bulk crystals, whose Fermi surfaces are composed of  both electron and hole pockets \cite{BaK_HDing, Terashima, BaK_YZhang}.  Built on the spin fluctuations between the electron and hole Fermi pockets,  a sign-changing $s$-wave pairing was predicted by  weak coupling theories \cite{Mazin, Kuroki, scalapino}. The second family consists of $A_x$Fe$_{2-y}$Se$_2$ (\textit{A}=K, Rb, Cs, Tl/K) \cite{YZhang, KFeSe_Ding, XJZhou, mou},  (Li$_{0.8}$Fe$_{0.2}$)OHFeSe \cite{XFLu, XHNiu}, the amonia-intercalated Li$_x$(NH$_2$)$_y$(NH$_3$)$_{1-y}$Fe$_2$Se$_2$ \cite{NH3}, and the recently discovered monolayer FeSe thin film \cite{qkxue, SYTan, RPeng, RPengN}.
The Fermi surfaces of this family of FeHTSs consist only of electron pockets. The nodeless gap observed in these iron-chalcogenides has posed a serious challenge to the weak coupling theories, which have predicted gap nodes for such a Fermi surface topology \cite{yao-li-wang}.   Therefore, a comprehensive understanding of  microscopic behaviors of both families is  critical for understanding the mechanism of FeHTS.
 
 Recently, a systematic  angle-resolved photoemission spectroscopy (ARPES)  investigation was conducted in many series of the first families of FeHTSs, including bulk Fe(Te,Se), BaFe$_2$(As$_{1-x}$P$_x$)$_2$, Ba(Fe$_{1-x}$Ru$_x$)$_2$As$_2$,  NaFe$_{1-x}$Co$_{x}$As and LiFe$_{1-x}$Co$_{x}$As \cite{ZRYe}. In all these compounds, the bandwidths are found to increase with the doping level $x$, as the superconductivity is weakened and eventually suppressed at high doping levels, regardless of the doping traits being heterovalent or isovalent. Such an observation is consistent with strong coupling theories, where pairing is mediated by local antiferromagnetic interactions \cite{dung-hai-davis, hu-hao, hu, hu-ma-lin, hu-ding, seo-hu, fang}. 
 It is then intriguing to ask if similar behavior is also present for the second family of FeHTS.
 However,
in contrast to the first  family of FeHTS,  an in-depth understanding of the doping dependence of  the second family of FeHTS is still lacking for several reasons.  The doping of many members of  the second family is discrete or cannot be controlled. In particular, for the 122$^{*}$ series of iron-chalcogenides ($A_x$Fe$_{2-y}$Se$_2$ and the like) that comprise most of this family, there is an intrinsic phase separation between a $\sqrt{5}$ $\times\sqrt{5}$ Fe-vacancy ordered insulating phase (also called the ``245" phase) and a superconducting phase \cite{mhfang1, mhfang2, FChenPhase, JZhao, tem}. Consequently, the doping-dependence of  the microscopic properties  of this series has not been investigated extensively. 

\begin{figure}
\includegraphics[width=8.7cm]{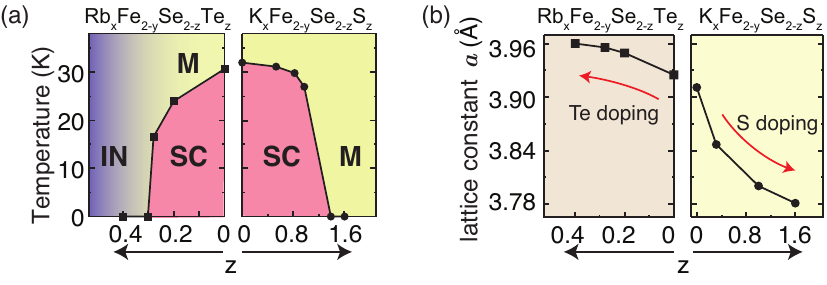}
\caption{(a) Phase diagrams of Rb$_x$Fe$_{2-y}$Se$_{2-z}$Te$_z$ and K$_x$Fe$_{2-y}$Se$_{2-z}$S$_z$. SC, M, and IN represent the superconducting, metallic and insulating phases, respectively. (b) In-plane lattice constants (data are extracted from Ref. \cite{LLSun, HCLei}) of Rb$_x$Fe$_{2-y}$Se$_{2-z}$Te$_z$ and K$_x$Fe$_{2-y}$Se$_{2-z}$S$_z$.} 
\label{transport}
\end{figure}

Recent  progress in materials synthesis has enabled the tuning of the superconductivity of $A_x$Fe$_{2-y}$Se$_2$ in a more controlled manner. As shown in Fig.~\ref{transport}(a), phase diagrams have been obtained when replacing Se with Te or S in $A_x$Fe$_{2-y}$Se$_2$ (\textit{A} = K, Rb) \cite{LLSun, HCLei}. Intriguingly, though the chemical pressure is tensile for Te doping and compressional for S doping as shown by their lattice constant evolution in Fig.~\ref{transport}(b), superconductivity is gradually suppressed in both cases. See the Supplementary Material  for in-plane resistivity and magnetic susceptibility data \cite{supplementary}. It is important to note that at low temperature the resistivity on the S-overdoped side is metallic, while that on the Te-overdoped side is insulating.  

In this work, we systematically studied the low energy electronic structure of  $A_x$Fe$_{2-y}$(Se, S/Te)$_2$ (\textit{A}=K, Rb, Tl/K), and  we show that they can be fit into one  generic phase diagram, with bandwidth or  correlation strength being the control parameter.
We found that on decreasing bandwidth, the ground state of these compounds evolves from a metal to a superconductor and then to  an insulator, which  has almost all the essential ingredients of the cuprate phase diagram, except that  bandwidth, rather than carrier doping,  is the control parameter. 
 Intriguingly, the   insulating phase found here has a  small  band filling and is  induced by band narrowing from a superconducting phase, which resembles the bandwidth-controlled Mott transition, except that the latter requires an {\slshape integer} band filling. Therefore it might  indicate either a novel correlation-driven insulator, or that the unit cell has been somehow expanded by certain ordering. 
Our results give a microscopic and unified understanding of various chemical-substitution-based phase diagrams of  the  122$^{*}$ series of iron-chalcogenide FeHTS. The revealed relationship between the superconductivity and the electronic correlations is reminiscent of similar observations made in iron pnictides \cite{ZRYe}, and thus might lead to a comprehensive understanding of both families of FeHTS.

\section{II. Experiment}
High-quality K$_x$Fe$_{2-y}$Se$_{2-z}$S$_z$ (\textit{z}~=~0, 0.53, 0.83, 0.97, 1.40, and 1.61) single crystals were synthesized using the self-flux method. High-purity Fe, Se, S and K were carefully mixed with the nominal compositions, then sealed in an evacuated quartz tube. The tube was heated to 1303~K and kept for 6~h, and then slowly cooled down to 1003~K at the rate of 3~K/h before shutting off the power. The S concentrations, \textit{z}, were confirmed by electron-probe micro-analysis (EPMA). Further details about the Rb$_x$Fe$_{2-y}$Se$_{2-z}$Te$_z$ (\textit{z}~=~0, 0.20, 0.28, 0.3, and 0.4) and Tl$_{0.4}$K$_{0.4}$Fe$_{1.7}$Se$_2$  single crystals can be found in Refs. \cite{LLSun, CJZhang}. Tl$_x$Fe$_{2-y}$Se$_{0.4}$S$_{1.6}$ and Tl$_x$Fe$_{2-y}$S$_{2}$ are non-superconducting compounds \cite{TlS}. High-resolution ARPES measurements were performed at the I05 beamline of the Diamond Light Source, beamline 5-4 of the Stanford Synchrotron Radiation Light source (SSRL), One-Cubed ARPES at BESSY~II and our in-house ARPES system with a Helium discharge lamp (21.2~eV photons), all equipped with a Scienta R4000 electron analyzer, and at beamline 28A of the Photon Factory, KEK, equipped with a Scienta SES-2002 electron analyzer. The overall energy resolution was 2.5$\sim$20~meV, depending on the photon energies and experimental setups, and the angular resolution was 0.3 degrees. Samples were cleaved under ultra-high vacuum conditions. Rb$_x$Fe$_{2-y}$Se$_{1.6}$Te$_{0.4}$ samples are measured at T=100~K to avoid charging effects and other samples are measured at T = 1$\sim$13~K depending on the experimental setups. The ARPES measurements on each sample were carried out within 8 hours, with sample aging effects carefully monitored.

\begin{figure}[t]
\includegraphics[width=8.7cm]{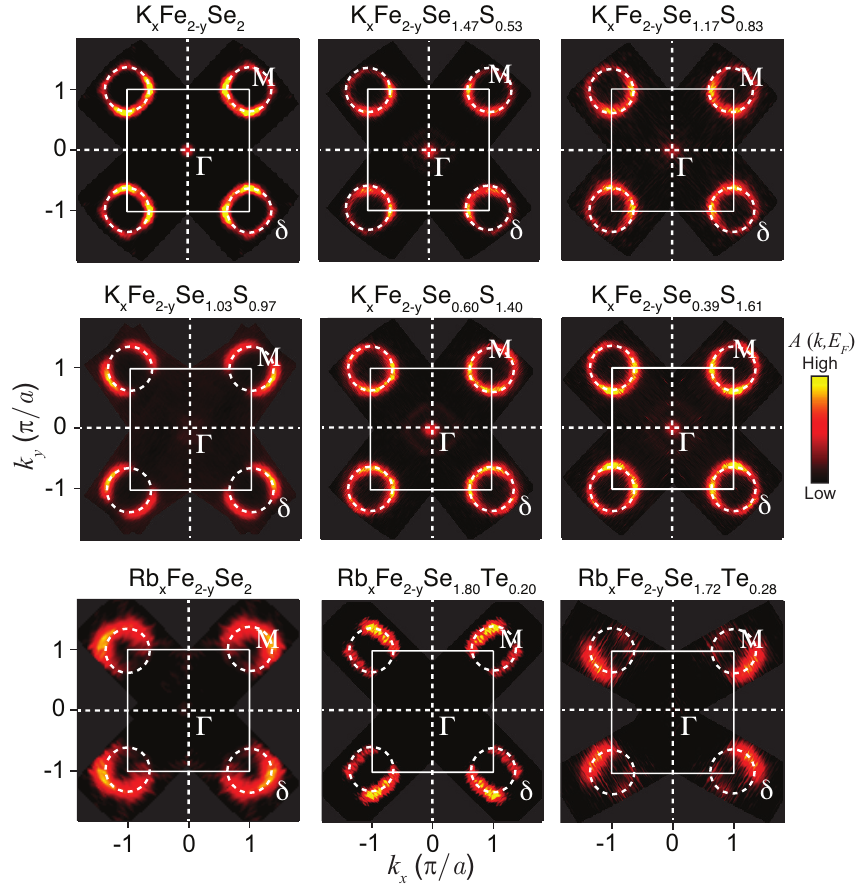}
\caption{Doping dependence of the Fermi surface sheets of Te/S doped $A_x$Fe$_{2-y}$Se$_{2}$. These nine false-color plots are the four-fold-symmetrized photoemission intensity maps at the Fermi energy ($E_F$) in the $\varGamma$-$M$ plane. The photoemission intensity distribution is mirror symmetrized with respect to the $k_x$ and $k_y$ axes. Intensities were integrated over a window of ($E_F$ $-$ 15~meV, $E_F$ + 15~meV). White dashed circles show the $\delta$ band Fermi crossings, wherever identifiable. } \label{mapping}
\end{figure}

\begin{figure*}
\includegraphics[width=18cm]{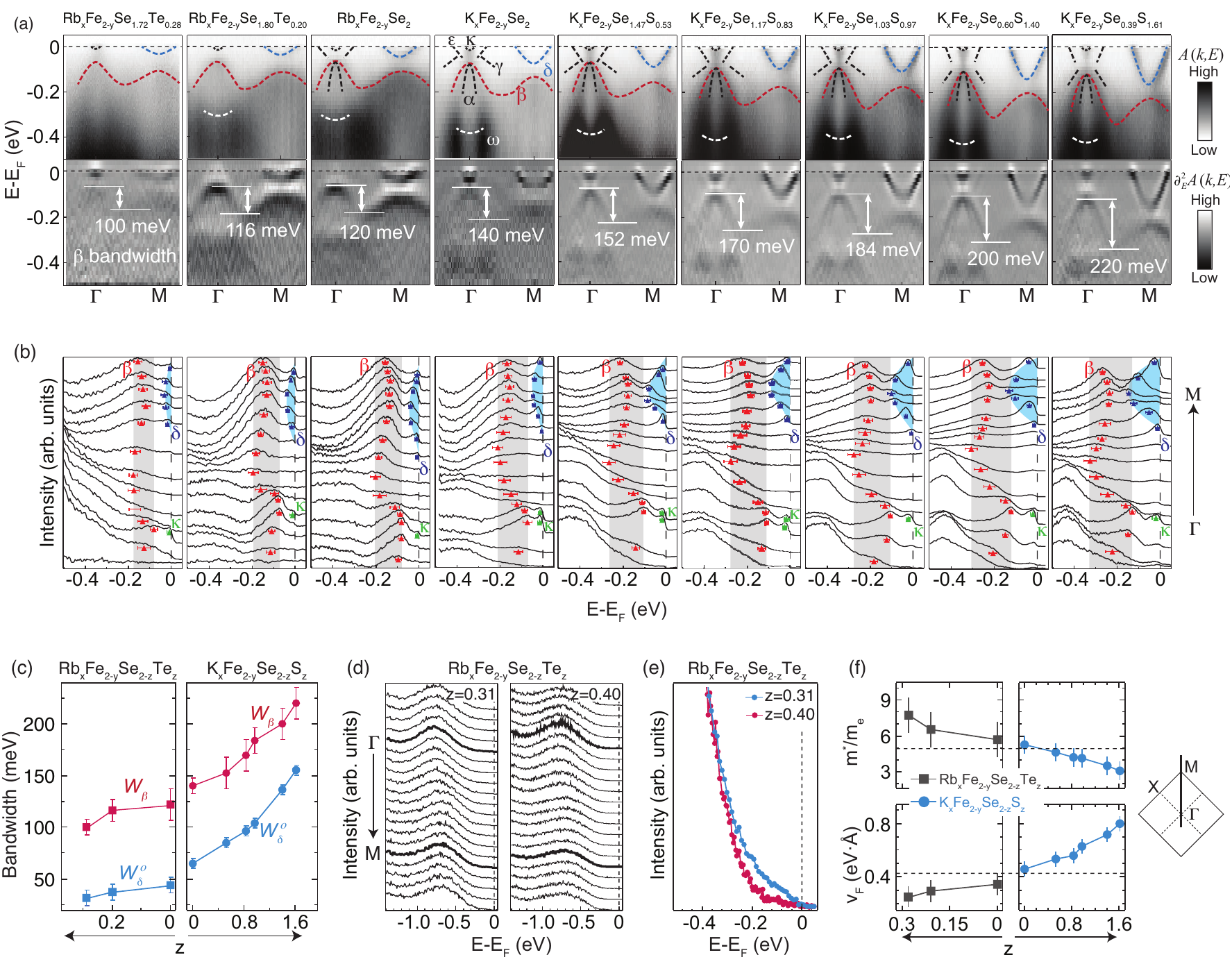}
\caption{(a) Doping dependence of the photoemission intensity distributions along the $\varGamma$-$M$ direction in Rb$_x$Fe$_{2-y}$Se$_{2-z}$Te$_z$ and K$_x$Fe$_{2-y}$Se$_{2-z}$S$_z$, and the corresponding second derivatives with respect to energy. The band structures were determined for the superconducting phases of these compounds \cite{FChenOrb}. The white double-headed arrows indicate the bandwidth of the $\beta$ band. 
(b) The background-subtracted EDCs for the data in (a).  The dispersions of the $\beta$ and $\delta$ band are clearly resolved, and the (occupied) bandwidth is highlighted by shadow.
(c) The $\beta$ bandwidth ($W_{\beta}$) and the occupied width of the $\delta$ band ($W_{\delta}^{O}$) as a function of doping in Rb$_x$Fe$_{2-y}$Se$_{2-z}$Te$_z$ and K$_x$Fe$_{2-y}$Se$_{2-z}$S$_z$. The error bars come from the uncertainty in the dispersion determination and curve fitting.  (d) The photoemission spectra of Rb$_x$Fe$_{2-y}$Se$_{2-z}$Te$_z$ for $z$~=~0.3 and $z$~=~0.4, along the $\varGamma$-$M$ cut marked on the sketch of the Brillouin zone in the inset on the right. (e) The comparison of the spectra taken at the $M$ point for $z$~=~0.3 and $z$~=~0.4 samples. (f) Doping dependence of the effective mass ($m^*$) and Fermi velocity ($v_F$) of the $\delta$ band. The $m^*$ and $v_F$ of Tl$_{0.4}$K$_{0.4}$Fe$_{1.7}$Se$_2$ are indicated by dashed lines. }
\label{bandwidth}  
\end{figure*}

\begin{figure}[t]
\includegraphics[width=8.7cm]{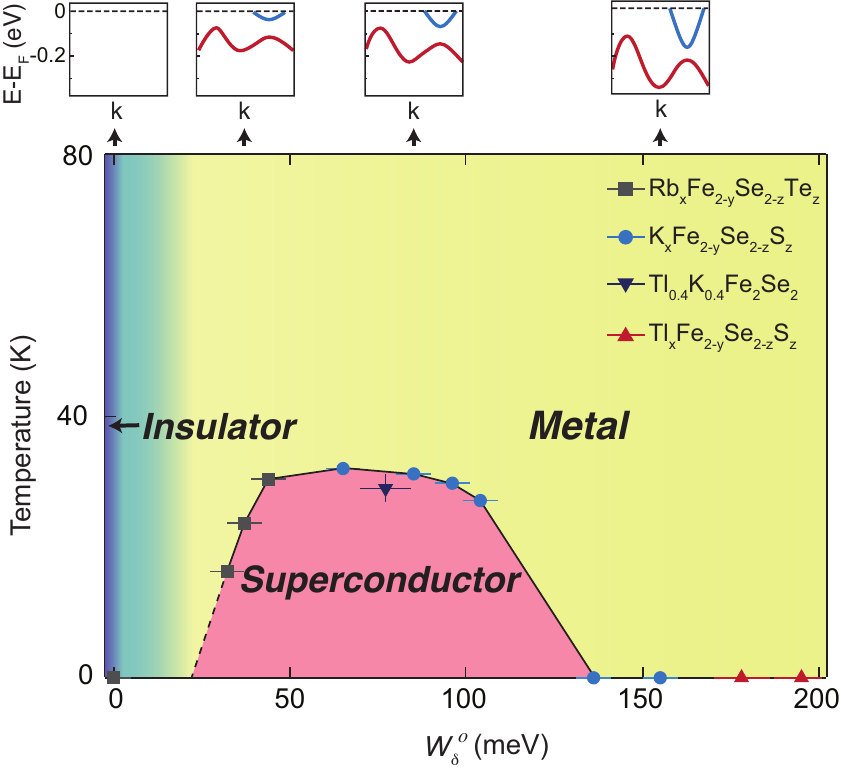}
\caption{A unified phase diagram of $T_c$ {\it vs}.\ $W_{\delta}^{O}$ for $A_x$Fe$_{2-y}$(Se,S,Te)$_2$ (\textit{A}=K, Rb, Tl/K). The sketches of the band dispersions for several dopings illustrate the band structure evolution and the absence of quasiparticles in the insulator phase. The dashed line is a linear extrapolation of the $T_c$ {\it vs}.\ $W_{\delta}^{O}$ data.
Note that the bandwidths for (Tl,K)$_x$Fe$_{2-y}$(Se,S)$_{2}$ are extroplated to match the doping level of the other materials \cite{supplementary}. }
\label{phasediag}
\end{figure}

\begin{figure}[t]
\includegraphics[width=8.7cm]{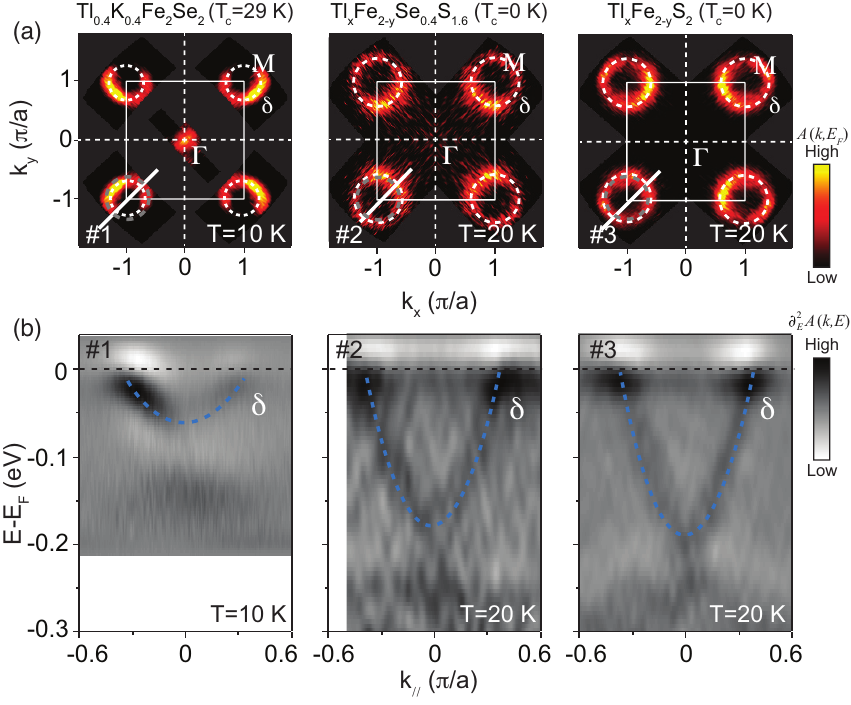}
\caption{ARPES data on Tl$_{0.4}$K$_{0.4}$Fe$_{1.7}$Se$_2$, Tl$_x$Fe$_{2-y}$Se$_{0.4}$S$_{1.6}$ and Tl$_x$Fe$_{2-y}$S$_{2}$. (a) False-color plots of the four-fold symmetrized photoemission intensity maps. Intensities were integrated over a window of ($E_F$-15~meV, $E_F$+15~meV). White dashed circles show the $\delta$ Fermi crossings, while the electron pockets of K$_x$Fe$_{2-y}$Se$_{2}$ are illustrated as dashed grey circles. (b) Photoemission intensity distributions at the zone corners along the cuts illustrated by white solid lines in (a).}
\label{tl}
\end{figure}

\section{III. Results}

We first examine the Fermi surface topologies of K$_x$Fe$_{2-y}$Se$_{2-z}$S$_z$ and  Rb$_x$Fe$_{2-y}$Se$_{2-z}$Te$_z$ in Fig.~\ref{mapping}. Similar to K$_x$Fe$_{2-y}$Se$_{2}$ \cite{YZhang, KFeSe_Ding, XJZhou}, only electron pockets can be observed at the zone center $\varGamma$ and the zone corner $M$. There are two electron-like bands $\delta$ and $\delta'$ around the $M$ point, which give nearly the same Fermi surface sheets \cite{FChenOrb, YMin}. However, only the shallower $\delta$ band can be clearly resolved [see Figure.~\ref{bandwidth}(a)] in our experimental geometry as reported before \cite{YZhang}. Around the zone center $\varGamma$, the shallow electron-like band $\kappa$ contributes a small amount of photoemission intensity, with a strong $k_z$ dispersion as observed before in K$_x$Fe$_{2-y}$Se$_{2}$ \cite{YZhang}. We present the $k_z$ dispersions of the $\delta$ and $\kappa$ Fermi surface sheets of K$_x$Fe$_{2-y}$Se$_{2-z}$S$_z$ (\textit{z}=0.53, 0.97) and Rb$_x$Fe$_{2-y}$Se$_{2}$ in Ref. \cite{supplementary}, which are independent of the S/Te concentrations. The weak $\varepsilon$ band in Figure.~\ref{bandwidth}(a) arises from the folding of the $\delta$ band by $\sqrt{2}$ $\times\sqrt{2}$ superstructures as reported before \cite{XJZhou, JZhao, FChenOrb, YMin, KFeSe-stm, liwei-stm}. 

Compared to the $\varepsilon$ band, the $\delta$ and $\delta'$ bands contribute most of the density of states (DOS) near the Fermi energy ($E_F$) and should dominate the low-energy electronic structure. Since both experiments and calculations found negligible $k_z$ dispersion for $\delta$ and $\delta'$ \cite{YZhang, FScal, supplementary}, we estimate the electron doping level based on the Fermi surface topologies in Fig.~\ref{mapping}, which gives 0.20 $\pm$ 0.02 electrons per Fe for both K$_x$Fe$_{2-y}$Se$_{2-z}$S$_z$ and Rb$_x$Fe$_{2-y}$Se$_{2-z}$Te$_z$ (assuming there is no iron vacancy in the superconducting phase \cite{KFeSe-stm, Chenxi}), independent of S/Te dopings. This finding indicates that S/Te doping does not introduce extra carriers into $A_x$Fe$_{2-y}$Se$_{2}$ (\textit{A} = K/Rb) within the error bar of our measurements. However, we found that the superconducting gap scales with $T_c$ for these compounds \cite{supplementary}. Furthermore, we did not see noticeable changes of the scattering rate with the increasing of S/Te concentration \cite{supplementary}. Thus the suppression of $T_c$ with the S/Te doping is not due to impurity effects. 

In contrast to the doping-independent Fermi surfaces and scattering rates, the S/Te doping clearly alters the band structure. Fig.~\ref{bandwidth}(a) shows the photoemission intensity distributions along the $\varGamma$-$M$ direction.  Here, the bandwidth of the $\beta$ band can be clearly resolved, and increases monotonically from left to right, in conjunction with the reduction of the in-plane lattice constant. The bottom of the $\delta$ band sinks accordingly. Such a band structure evolution is further illustrated by energy distribution curves (EDCs) in Fig.~\ref{bandwidth}(b), where a background has been subtracted to retrieve the dispersion more clearly as described in the  Supplementary Material \cite{supplementary}. 
 Figure~\ref{bandwidth}(c) summarizes the systematic evolution of the bandwidth of $\beta$ ($W_{\beta}$) and the occupied width of the $\delta$ band ($W_{\delta}^{O}$) with S/Te doping, where $W_{\beta}$  is enhanced by 120~\%, and $W_{\delta}^{O}$ by a factor of more than 4, on decreasing the lattice constants or correspondingly, increasing the chemical pressure.

Figure~\ref{bandwidth}(d) shows the photoemission spectra along $\varGamma$-$M$ for Rb$_x$Fe$_{2-y}$Se$_{2-z}$Te$_z$ (\textit{z}=0.3, 0.4). The spectra do not show any noticeable momentum dependence. There are no quasiparticles around the chemical potential, as expected for insulators. However, the \textit{z}=0.3 sample has some residual spectral weight near the chemical potential, while the \textit{z}=0.4 sample has none [Fig.~\ref{bandwidth}(e)]. This may due to the coexistence of some trace amount of superconducting phase in the \textit{z}=0.3 sample as indicated in the resistivity data \cite{LLSun}, or the incoherent remains of the suppressed quasiparticle bands. Moreover, our high-resolution transmission electron microscope (HRTEM) data of Rb$_x$Fe$_{2-y}$Se$_{1.6}$Te$_{0.4}$ show that the $\sqrt{5}$ $\times\sqrt{5}$ vacancy-ordered ``245'' insulating phase coexists with a $\sqrt{2}$ $\times\sqrt{2}$ superstructured phase in mesoscopic scale \cite{supplementary} .  Since a similar $\sqrt{2}$ $\times\sqrt{2}$ superstructure exists in the superconducting phase, as shown by the band folding in Fig.~\ref{bandwidth}(a), one can conclude that this phase with the $\sqrt{2}$ $\times\sqrt{2}$ superstructure evolves from a superconductor to an insulator with increasing Te doping. This insulator is distinct from that induced by Fe vacancies in the ``245''  phase \cite{FChenPhase, sqm1, sqm2}. Moreover, since Te only introduces chemical pressure instead of altering the band filling, such a sudden disappearance of Fermi surface when the bandwidth diminishes indicate a correlation-driven superconductor/metal-insulator transition.
 Due to the strong competitions among various inter-site exchange interactions and strong spin fluctuations in the iron chalcogenides, as exemplified in bulk FeSe \cite{FeSe-cava, cao-gong}, it is an open question whether there is a magnetic order in this  insulator phase with the $\sqrt{2}$ $\times\sqrt{2}$ superstructure.  This might be answered by future neutron scattering studies.

As shown in Ref. \cite{supplementary}, band calculations indicate that the bare bandwidths of KFe$_{2}$Te$_{2}$, KFe$_{2}$Se$_{2}$ and KFe$_{2}$S$_{2}$ only differ by about 63\% ($W_{\beta}$) and 19\% ($W_{\delta}^{O}$) \cite{FScal}. By applying a linear interpolation, we estimate that the change of the bare bandwidth in the measured doping range here should be within 32\% ($W_{\beta}$) and 10\% ($W_{\delta}^{O}$). The remarkable change of the bandwidths in Fig.~\ref{bandwidth} is thus due to the additional bandwidth renormalization by electronic correlations, which is driven largely by Hund's rule coupling in the FeHTSs \cite{yin}. Similar behaviors have been observed in the prototypical bandwidth-controlled Mott transition of NiS$_{2-x}$Se$_{x}$ \cite{Xu:2014em}. 
Because the bare bandwidth and the effective bandwidth are monotonically related, either of them can be used to characterize the itinerancy or kinetic energy. However,
since the effective bandwidth  not only represents the itinerancy of quasiparticles, but also carries the information on the electronic correlations in these compounds, we use the effective bandwidth   in the following discussions.

In addition to the bandwidth, effective mass ($m^*$) and Fermi velocity ($v_F$) of the $\delta$ band reflect the electronic correlation strength as well. As shown in Fig.~\ref{bandwidth}(f), $m^*$ decreases from the Te-overdoped side to the S-overdoped side, while $v_F$ shows the opposite trend accordingly. Taking $W_{\delta}^{O}$ as the electronic parameter, one can obtain the phase diagram in Fig.~\ref{phasediag}, which shows that superconductivity is suppressed by driving the bandwidth away from its optimal value by doping either the isovalent S or Te. Furthermore, we have examined whether other sibling iron-chalcogenides can fit into this phase diagram. Data for Tl$_{0.4}$K$_{0.4}$Fe$_{1.7}$Se$_2$  ($T_c$ = 29 K) and Tl$_x$Fe$_{2-y}$Se$_{2-z}$S$_z$ (non-superconducting) are shown in Figs.~\ref{tl}(a) and (b). Compared with K$_x$Fe$_{2-y}$Se$_{2}$, the electron Fermi pockets of Tl$_{0.4}$K$_{0.4}$Fe$_{1.7}$Se$_2$ are smaller [Fig
.~\ref{tl}(a)], corresponding to a carrier concentration of 0.14 $\pm$ 0.02 electrons per Fe. However, assuming these off-plane dopants do not affect the band renormalization, as found for   Ba$_{1-x}$K$_x$Fe$_2$As$_2$ before,  
 we rigidly shift the parabolic band so that the extrapolated Fermi crossings match those of K$_x$Fe$_{2-y}$Se$_{2}$ \cite{supplementary}, the extrapolated $W_{\delta}^{O}$ would fall into the curve shown in Fig.~\ref{phasediag} within the error bar. For the non-superconducting Tl$_x$Fe$_{2-y}$Se$_{0.4}$S$_{1.6}$ and Tl$_x$Fe$_{2-y}$Se$_{2}$, their electron pockets are slightly larger than those of K$_x$Fe$_{2-y}$Se$_{2}$, while their $\delta$ bands are much broader (Fig.~\ref{tl}). After compensating the small difference in the Fermi pockets by rigid shifting their Fermi crossings to match those of K$_x$Fe$_{2-y}$Se$_{2}$, one can obtain their $W_{\delta}^{O}$s. They are still much larger than 150 meV, far above the superconducting regime in Fig.~\ref{phasediag}.  We note that a similar phase diagram  can be reached if using $m^*$ as the horizontal axis.

\section{IV. Discussions}
 
We have observed the bandwidth evolution of the 122$^{*}$ series of iron-chalcogenides
by substituting Se with S or Te,  and obtain a unified phase diagram in Fig.~\ref{phasediag} from metal (S side) to superconductor, then to an insulator (Te side), with nearly fixed Fermi surface  and negligible impurity scattering effects (see Supplementary Materials \cite{supplementary}).  Moreover,  there is no noticeable orbital occupation change either.
The dramatic change of bandwidth (or band bottom of $\delta$) signifies strong correlation effects in iron-selenides. Such a bandwidth-controlled or correlation-driven superconductor-to-insulator transition resembles the bandwidth-controlled  or the Brinkman-Rice Mott transition, where  the bandwidth decreases to zero or the quasiparticle mass diverges  \cite{brinkman-rice}.
Interestingly, a Mott transition, no matter in a single-band or multiple-band system, requires integer filling of the band \cite{brinkman-rice, Takuya}, whereas there is a small band filling in the insulating state of Te-overdoped sample. Therefore, it is not a  Mott insulator or at least not a usual one, instead, it could be a novel correlation-driven insulator  where Anderson localization might play a role. Alternatively, judging from the depleted spectral weight at the chemical potential, it is also likely that certain order, such as charge density wave, might have occured and multiplied the unit cell, which in turn would cause an integer band filling, and give a Mott transition.  
Recently, J. He and coauthors reported a carrier-doping-induced insulating phase in monolayer FeSe  \cite{JHe_2014}. However, it was later found by Y. Fang and coauthors that similar band structure and insulating phase could be reproduced in FeSe films with iron vacancies \cite{SYTan_2015}. Therefore, the insulating phase in FeSe film is qualitatively different from the insulating Rb$_x$Fe$_{2-y}$Se$_{2-z}$Te$_z$ (\textit{z}=0.3, 0.4).

In Fig.~\ref{phasediag}, there is a metallic regime between the insulating phase and superconducting phase. If this regime existed in real materials, it would correspond to the narrow tellurium concentration range of 0.28 $<$ $\textit{z}$ $<$ 0.3. Since this doping range is quite narrow, the sudden change between insulating state and superconducting state indicates that the superconductor-to-insulator transition here is likely of first order, and this metallic regime may not exist in real materials. Consistently in this narrow doping range, the transport behaviors evolve drastically, and there is some trace of the superconducting phase, an indicative of phase separation \cite{LLSun}.

Towards the large bandwidth or weak correlation regime, the behavior of the  122$^{*}$ series of iron-chalcogenides  resembles the observations made before in the first family of FeHTS  \cite{ZRYe}. That is, the superconductivity is weakened and eventually suppressed at the high doping level with increased bandwidth. 
 However, in the strong correlation regime, most iron pnictides enters the competing collinear antiferromagnetic metallic ground state instead of an insulating state found here. Nevertheless, all these results jointly suggest that the bandwidth or correlation strength provides a unifying standpoint in understanding the ground-state evolutions in both families of FeHTS’s.

Like the $t$-$J$ model for the cuprate, FeHTSs have been described by effective models with short-range antiferromagnetic exchange couplings in strong local pairing scenarios, which  could give  the pairing symmetries and gap functions   for both families of FeHTSs \cite{dung-hai-davis, hu-hao, hu, hu-ma-lin, hu-ding, seo-hu, fang}.
In such effective models, there are two critical parameters that control the superconductivity: the renormalized effective bandwidth, $W$, and the effective antiferromagnetic coupling strength $J$ (or multiple $J_i$s among various neighbors). The superconductivity appears when the ratio of $J/W$ is bound within a certain region. Below the lower bound, electron correlations are too weak, so that the system cannot sustain the unconventional high temperature superconductivity; while beyond the upper bound, too strong electron correlations or too weak itinerancy would give an insulator and/or magnetic ordered phase instead of a superconductor. If the effective short-range antiferromagnetic exchange couplings $J$s do not vary significantly with the isovalent doping, as in the case of FeHTSs \cite{Dai:2012em, Dai:2013}, the bandwidth (or other equivalent parameters describing  electron itinerancy, or conversely, correlation) becomes  the pivotal variable that controls  the evolution of superconductivity, as observed here   in these 122$^{*}$ series of iron-chalcogenide superconductors. 
 Therefore, the observed robust correlation between the superconductivity and $W$ provides a compelling support for the local pairing scenario in the 122$^{*}$ series of iron-chalcogenides superconductors.
Furthermore, in the strong coupling theories that tried to understand both FeHTSs  and cuprates \cite{dung-hai-davis, hu-ding}, the phase evolution is controlled by the competition between the kinetic energy and interactions. Therefore, it is not surprising that the phase diagram found here resembles the cuprate phase diagram, except that the control parameter here is bandwidth, instead of carrier doping.  Note that in the cuprate case,  with increased doping,  the effective bandwidth increases, and the correlation strength decreases. As a result, it is possible to understand both the FeHTSs  and the cuprates in such a unified way.

Finally, we note that our data further challenges the weak coupling mechanism, which stresses the interactions of electrons around the Fermi surface. In this kind of scenarios, the stronger correlations (or narrower bandwidth, smaller $v_F$) give rise to a larger DOS, which is expected to yield higher $T_c$, when the Fermi surface is invariant upon isovalent doping. However, our data suggests that the superconductivity will be strongly suppressed once the electron correlation strength goes beyond the optimal regime, even though the DOS continuously increases.  For example, Rb$_x$Fe$_{2-y}$Se$_{2-z}$Te$_z$ has stronger correlations but a lower $T_c$ than Rb$_x$Fe$_{2-y}$Se$_{2}$. Moreover, Tl$_{0.4}$K$_{0.4}$Fe$_{1.7}$Se$_2$  and K$_x$Fe$_{2-y}$Se$_{2}$ have almost the same $T_c$, $m^*$ and $v_F$ [dashed line in Fig.~\ref{bandwidth}(f)], while there is significant difference between their sizes of Fermi surfaces.  This is not consistent with the weak coupling mechanism as well.

\section{V. Conclusions}

To conclude, we have shown that the superconductivity evolutions in K$_x$Fe$_{2-y}$Se$_{2-z}$S$_z$,  Rb$_x$Fe$_{2-y}$Se$_{2-z}$Te$_z$ and (Tl,K)$_{x}$Fe$_{2-y}$Se$_{2-z}$S$_{z}$  can be comprehensively understood from their bandwidth evolutions.  We have found a correlation driven phase evolution from metal to superconductor, and eventually to a non-trivial insulating phase.  Our data suggest that correlation (or bandwidth as the electronic control parameter) plays a critical role in the iron based superconductors with only electron Fermi surfaces, as found before for those with both electron and hole Fermi surfaces. Our results thus   present an important step to the unified understanding of  all the iron-based superconductors, which are consistent with  the strong local pairing scenarios.

\section{Acknowledgments}
We gratefully acknowledge Dr. M. Hoesch, Dr. T. Kim, and Dr. P. Dudin at Diamond, Dr. K. Ono at KEK, Dr. D. H. Lu and Dr. H. Makoto at SSRL, and Dr. E. Rienks at BESSY II for ARPES experimental support, and we appreciate Prof. R. C. Che and Dr. Q. S. Wu at Fudan University for their great help in the TEM measurements. This work is supported by the National Science Foundation of China and National Basic Research Program of China (973 Program) under Grants No. 2012CB921402, No. 2012CB927401, No. 2011CBA00106, No. 2011CBA00103, No. 2012CB821404, and No. 2015CB921004. SSRL is operated by the US DOE Office of Basic Energy Science.

\end{document}